# How Money Enters Before It Leaves: Experimental Remuneration and the Mobile-Payment Effect

**Yizhao Jiang**

School of International Development and Cooperation, University of International Business and Economics, Beijing, China

Corresponding author: Yizhao Jiang, No. 10 Huixin Dongjie, Chaoyang District, Beijing 100029, China; email: yizhao.jiang@uibe.edu.cn

**Abstract**

Consequential purchasing experiments typically focus on how money leaves a consumer. When an experiment provides spendable funds before valuation, however, the preceding inflow can also shape the payment contrast subsequently measured. In a rule-assigned 2 × 2 experiment, university participants received an RMB 30 participation payment either in cash or by WeChat transfer and completed any binding purchase either in cash or through WeChat Pay. They reported willingness to pay for two displayed products in an incentivized BDM-style procedure. The primary sample contains 522 protocol-valid integer valuations from 274 participants across two implementations. Mobile rather than cash payment changed adjusted willingness to pay by −RMB 0.39 following cash remuneration and by +RMB 2.21 following mobile remuneration. The difference between these conditional contrasts was RMB 2.60 (95% CI [0.86, 4.34]), or 0.60 residual standard deviations; a participant-level, within-wave permutation test yielded $p = .004$. The interaction was similar for a mug and a face-value voucher pair and directionally consistent across implementations. Exploratory decomposition locates the asymmetry in lower valuation when mobile-transferred remuneration is followed by cash payment. The findings document short-run receipt–payment sequence dependence and show that experimental remuneration forms part of the payment environment: an estimate attributed to a payment medium can depend on how immediately available funds first entered the consumer's budget.

## 1. Introduction

Payment research usually begins at checkout: does a consumer value or buy more when payment is made digitally rather than in cash? Consequential experiments add an earlier event that is easy to overlook. To make a real transaction feasible, the experiment may first place spendable money in the participant's hands or account. The resulting choice then contains two linked monetary events—how money enters and how it leaves. If the participation payment is behaviorally neutral, the estimated mobile-payment contrast should be invariant to the medium in which that money was received.

That invariance is not guaranteed. Classic studies associate card payment with greater spending or willingness to pay (WTP), but later evidence shows that cashless effects are modest, heterogeneous, and increasingly contingent on context (Prelec & Simester, 2001; Liu & Dewitte, 2021; Schomburgk et al., 2024). Payment transparency, payment coupling, price attention, and the pain of paying can change the behavioral meaning of an outflow. Field evidence likewise indicates that shifts toward digital payment can raise actual consumption, with effects that depend on the transaction environment (Agarwal et al., 2024). A payment medium is therefore not a stable treatment that can be separated automatically from the budget in which it operates.

The route by which money enters that budget may matter as well. Windfalls, labels, and acquisition-related feelings can direct consumption, while economically equivalent funds may remain partitioned by source, purpose, or payment account (Thaler & Johnson, 1990; Levav & McGraw, 2009; Milkman & Beshears, 2009). Payment cards and mobile-money transfers can themselves form behaviorally meaningful account boundaries (Gelman & Roussanov, 2024; Lee et al., 2024). Experimental funds consequently have an origin and a form as well as an amount.

Existing payment studies do not isolate this possibility. Runnemark et al. (2015) compared three combinations of remuneration and purchase media but omitted account remuneration followed by cash purchase because participants might lack cash. Liu et al. (2021) varied the source of money while aligning the medium in which it was provided with the medium used for payment. More generally, recent experiments demonstrate that incentive protocols can affect elicited preferences (Aydogan et al., 2024; Berlin et al., 2026), but they do not ask whether the medium of remuneration changes a cash–mobile payment contrast. The missing design is a complete crossing of the medium through which money arrives with the medium through which it is spent.

We provide that design. Before valuation, participants received RMB 30 either as cash or by WeChat transfer. Any binding purchase was then completed either in cash or through WeChat Pay, producing four receipt–payment paths: cash–cash, cash–mobile, mobile–cash, and mobile–mobile. Participants were instructed to bring sufficient funds in both media and confirmed their ability to pay up to RMB 20, making every assigned path feasible. They valued a physically displayed mug and a pair of McDonald's vouchers using an incentivized Becker–DeGroot–Marschak-style procedure. The experiment was implemented at two Chinese universities in December 2018 and June 2019, a dual-medium period in which WeChat Pay was routine while cash remained familiar and usable. The mobile-remuneration/cash-payment condition is more than an additional treatment cell: it identifies whether the cash–mobile contrast is invariant to how experimental funds enter the decision environment.

The design makes remuneration-medium neutrality a direct empirical benchmark. Let $\Delta C$ denote the mobile-versus-cash payment contrast after cash remuneration and $\Delta M$ the corresponding contrast after mobile remuneration. Neutrality requires $\Delta M - \Delta C = 0$. The data reject that restriction. The adjusted payment contrast is −RMB 0.39 after cash remuneration and +RMB 2.21 after mobile remuneration, yielding an interaction of +RMB 2.60 (95% CI [0.86, 4.34]). The estimate is about 0.60 residual standard deviations, survives participant-level within-wave permutation inference, and is similar across the two products and directionally consistent across the two implementations. The four-cell pattern is asymmetric: valuation is lowest when mobile-transferred remuneration is followed by cash payment.

The study makes two contributions. First, it identifies an overlooked design condition in payment experiments. When a study supplies spendable funds before valuation, the remuneration medium can alter the contrast attributed to the purchase-payment medium. The relevant estimand is therefore a conditional contrast embedded in a receipt–payment path, rather than a single cash–mobile coefficient detached from the way experimental funds entered the decision environment.

Second, it connects payment behavior to the organization of newly available money. We use short-run receipt–payment path dependence to describe the empirical regularity that valuation depends on the sequence formed by the medium of a salient recent inflow and the medium required for the ensuing outflow. This pattern identifies a behaviorally meaningful boundary consistent with incomplete fungibility across receipt and payment environments. Medium-linked mental accounts, account salience, cross-medium action friction, and the heightened salience of relinquishing physical cash offer plausible interpretations.

## 2. Conceptual Background

### *2.1 Payment-medium effects are conditional*

In a frictionless benchmark, payment media are interchangeable implementations of the same budget constraint. Behavioral evidence rejects that invariance. Payment cues can facilitate spending (Feinberg, 1986), payment method can change WTP (Prelec & Simester, 2001), and payment mechanisms can alter the rehearsal, immediacy, and memorability of expenditure (Soman, 2001). Payment form also affects how transparently a financial outflow is represented and how tightly consumption is coupled to its cost (Prelec & Loewenstein, 1998; Soman, 2003; Raghubir & Srivastava, 2008).

The pain of paying is a prominent account of these effects, but it is neither unitary nor mechanically linked to spending. Transaction-state discomfort, stable tightwad–spendthrift tendencies, payment transparency, loss salience, price attention, and transaction fluency are related but distinct constructs (Rick et al., 2008; Reshadi & Fitzgerald, 2023). Digital payment may reduce the visibility of an outflow, increase attention to consumption rewards, or lower transaction friction (Ma et al., 2024). A difference in perceived pain therefore need not translate one-for-one into a difference in valuation.

The evidence accordingly points to a conditional rather than universal payment effect. Debit-card payment can increase WTP even when account depletion is immediate (Runnemark et al., 2015), yet the classic card effect and its extension to mobile payment are not consistently reproduced (Liu & Dewitte, 2021). Meta-analytic evidence finds a small positive average cashless effect accompanied by contextual heterogeneity and temporal attenuation (Schomburgk et al., 2024). Survey evidence indicates that electronic payments can feel less painful than cash (Broekhoff & van der Cruijsen, 2024), while large-scale field evidence shows that an externally induced transition toward digital payment can increase actual consumption (Agarwal et al., 2024). Taken together, this literature makes the payment environment—not payment technology alone—the appropriate unit of analysis.

### *2.2 Experimental remuneration and the mental organization of money*

Experimental remuneration is designed to make choices consequential, but it also creates an economically and psychologically salient inflow. The medium, incidence, and probability of experimental payment are substantive design features rather than purely logistical details (Clot et al., 2018; Voslinsky & Azar, 2021). Recent experiments further show that paying all participants, only a subset, or none can change elicited risk and time preferences (Aydogan et al., 2024; Berlin et al., 2026). Incentives therefore have both a delivery protocol and a representational form.

Mental accounting provides the broader framework for why the form and source of an inflow can survive into later choice. Consumers organize money by source, purpose, and use, creating psychological budgets whose boundaries need not coincide with the aggregate budget constraint (Thaler, 1985, 1999; Henderson & Peterson, 1992; Heath & Soll, 1996). Within this framework, prior gains alter risk taking (Thaler & Johnson, 1990), unexpected windfalls may be spent more readily than other wealth (Arkes et al., 1994), and small windfalls can be allocated through distinct mental accounts even in ordinary shopping (Milkman & Beshears, 2009). The source and perceived ownership of an endowment also change behavior (Aycinena et al., 2026). A participation payment received immediately before valuation is therefore not necessarily absorbed without residue into undifferentiated wealth.

Labels, restrictions, and category-specific budgets further limit fungibility. Acquisition-related feelings can label money and shape what it is used to purchase (Levav & McGraw, 2009). Explicit labels redirect incentivized consumption (Abeler & Marklein, 2017), while restricted-use funds alter product choice (Reinholtz et al., 2015). Field evidence likewise shows that category-specific budgets can generate consumer responses inconsistent with full fungibility (Hastings & Shapiro, 2013).

Payment media can themselves become account boundaries. An exogenously received payment card can increase total expenditure when consumers spend from the new card without proportionately reducing expenditure on existing cards (Gelman & Roussanov, 2024). Conversely, mobile money associated with migrant remittances can be less readily spent than cash because it retains a source-specific social meaning (Lee et al., 2024). Digital money is thus neither inherently more nor less spendable. Its behavioral availability depends on how it enters the budget and on the payment environment in which the next choice occurs.

### 2.3 Remuneration-medium neutrality and the factorial estimand

Define the mobile-payment contrast after cash remuneration as:

$$\Delta C = E[WTP \mid ReceiveCash, PayMobile] - E[WTP \mid ReceiveCash, PayCash].$$

Define the corresponding contrast after mobile remuneration as:

$$\Delta M = E[WTP \mid ReceiveMobile, PayMobile] - E[WTP \mid ReceiveMobile, PayCash].$$

Remuneration-medium neutrality requires:

$$\Delta M - \Delta C = 0.$$

The subscripts denote the remuneration medium. This is a benchmark implied by the factorial design, not a directional claim about any single psychological process. It does not require the checkout medium itself to be neutral: a mobile-payment effect may exist, but its magnitude must remain unchanged when the same participation amount is delivered through a different medium.

The complete 2 × 2 design makes the restriction directly testable. The receipt-medium × payment-medium interaction estimates the difference between the two conditional payment contrasts. A nonzero interaction means that the measured mobile-payment contrast depends on the path through which immediately available money first entered the participant's budget. Because physical and digital balances can differ in how readily they are integrated with existing funds, a departure from neutrality need not be symmetric across receipt–payment paths. The four cell means reveal the empirical form of that dependence. We reserve directional explanations of the observed asymmetry for the Discussion, where they can be evaluated as interpretations rather than treated as additional primary hypotheses.

## 3. Method

### 3.1 Participants, implementations, and allocation

The experiment was conducted in two scheduled university sessions. The first took place on 12 December 2018 at the Baoding campus of North China Electric Power University; the second took place on 22 June 2019 at the Beijing Changping campus of China University of Petroleum. The pooled source files contain 275 participant records—155 in the first implementation and 120 in the second. Each implementation occurred at a different university and date; these features are therefore absorbed by a single implementation-wave indicator.

Participants were assigned by fixed intervals of the final two digits of their university-issued student numbers (00–24, 25–49, 50–74, and 75–99), mapped in advance to the four factorial cells. Published coding standards at both universities identify the terminal block as an administratively assigned serial number rather than a code for major, sex, or payment behavior.[1] The identifiers predated recruitment and were not selectable by participants or researchers. We therefore treat assignment as plausibly as-if random within implementation waves and complement participant-clustered regression with within-wave permutation inference.

The study protocol was approved by the Claremont Graduate University Institutional Review Board (CGU IRB #3320). Participants reviewed an informed-consent statement before taking part and provided consent to participate.

The sample size was determined by the two scheduled implementation sessions and the available participant pools; no prospective power calculation was recorded.

### 3.2 Full-factorial design

The experiment independently varied the medium of the participation payment and the medium required for a consequential purchase. Before reporting WTP, participants received RMB 30 either as cash or by WeChat transfer. Any purchase generated by the elicitation procedure was then completed either in cash or through WeChat Pay. The design therefore produced four receipt–payment paths, summarized in Table 1.

**Table 1. Full-factorial design and participant counts**

| Remuneration medium | Purchase-payment medium | Condition | Participants |
|---|---|---|---|
| Cash | Cash | Cash–cash | 77 |
| Cash | WeChat Pay | Cash–mobile | 58 |
| WeChat transfer | Cash | Mobile–cash | 71 |
| WeChat transfer | WeChat Pay | Mobile–mobile | 68 |

*Notes.* Participant counts include those contributing at least one protocol-valid integer WTP response.

Participants had been asked to bring sufficient cash and to maintain sufficient mobile-payment funds to cover the maximum possible purchase price. At the session, they confirmed that both media contained enough funds to complete a purchase of up to RMB 20. The procedure therefore addressed payment feasibility while preserving the experimental variation in remuneration medium. Immediately before valuation, participants assigned to cash payment placed the required cash on the desk, whereas those assigned to mobile payment opened the mobile-payment application. The complete treatment wording is reported in Online Appendix Table B.1.

### *3.3 Products and consequential WTP elicitation*

Participants valued a ceramic mug and two RMB 10 McDonald's vouchers with a combined nominal face value of RMB 20. Both products were physically displayed before bidding. The mug had an approximate retail reference price of RMB 20, but this information was not displayed or communicated to participants. The products were positioned on a broadly similar nominal monetary scale while differing in whether an explicit value anchor was available to participants.

WTP was elicited with a Becker–DeGroot–Marschak-style procedure (Becker et al., 1964). Participants submitted an integer bid from RMB 0 to RMB 20 for each product before learning which product would be implemented. The parity (odd or even) of the experimental questionnaire ID preassigned one product to be consequential, and that assignment was revealed only after both bids had been submitted. Each bid was therefore potentially consequential when reported. A binding integer price from RMB 0 to RMB 20 was generated on site with a random-number generator. If the participant's WTP for the consequential product was at least the generated price, the participant purchased it at that price using the assigned payment medium; otherwise no purchase occurred.

The instructions emphasized that the transaction price was generated independently of the bid. They included numerical examples and a comprehension question. Researchers reviewed the examples, revealed and explained the correct answer, and answered procedural questions before participants submitted WTP. Under the standard assumptions of the BDM mechanism, the independent random price and binding purchase made truthful bidding optimal. These procedures were intended to support participants' understanding of the elicitation rule.

The study was designed to measure valuation. Actual purchase incidence is not a secondary outcome with a common threshold: it is jointly determined by submitted WTP and an independently generated random price. We therefore analyze WTP rather than comparing raw purchase rates across conditions.

### *3.4 Questionnaire measures*

After the valuation task, participants completed measures of demographic characteristics, payment frequency and preferences, an adapted tightwad–spendthrift measure (Rick et al., 2008), and perceptions of mobile payment. The second implementation added BIS/BAS items (Carver & White, 1994) and a financial-monitoring scale. Because these measures differ across waves, and because the attitudinal measures were collected after valuation, they are not used as mediators of the primary effect. Common pre-existing characteristics—sex, income category, phone-use time, and reported historical use frequencies for cash, Alipay, and WeChat Pay—are used only in a descriptive covariate-adjustment robustness analysis. Questionnaire-based moderation analyses are designated exploratory and reported in the online appendix. Additional disclosed items include the Implementation 1 tablet choice, nonbinding intended quantities, a purchase-contingent repurchase WTA item, and condition- or implementation-specific mobile-account questions. These items are documented in the study materials and codebook but are not treated as manuscript outcomes.

### 3.5 Data construction and primary sample

The pooled long-format source file contains 550 product records from 275 participants. We reconstruct participant identifiers by combining implementation wave and the within-wave experimental questionnaire ID because the two source files reused some questionnaire ID values. This ID is distinct from the university student number used for treatment-version assignment. The primary sample excludes missing, out-of-range, and non-integer bids, yielding 522 protocol-valid product observations from 274 participants. The complete sample flow is reported in the online appendix.

We retain every protocol-valid product response rather than requiring both responses from each participant. Requiring paired completeness produces differential retention across the four cells and can therefore introduce treatment-related sample selection. Product-specific response-validity diagnostics are reported in the online appendix, and a paired-complete integer sample of 248 participants and 496 observations is used as a demanding robustness check. We also estimate participant-mean models so that participants contributing two bids receive no greater weight than those contributing one.

All participants received a guided explanation of the correct answer before bidding. Recorded comprehension status was not used as a primary exclusion criterion; robustness checks excluding recorded failures and retaining recorded passes only are reported in the online appendix.

### 3.6 Empirical strategy

The primary factorial model is:

$$WTP_{ij} = \alpha + \beta_1 \; PayMobile_i + \beta_2 \; ReceiveMobile_i + \beta_3 \; (PayMobile_i \times ReceiveMobile_i) + \gamma_j + \delta w + \varepsilon_{ij}.$$

where i indexes participants, j products, and w implementation waves. PayMobile equals one when the assigned purchase medium is WeChat Pay, and ReceiveMobile equals one when the participation payment is delivered by WeChat transfer. The preferred model includes product and implementation-wave fixed effects. Standard errors are clustered by participant to preserve the two-product repeated-measures structure.

The coefficient $\beta_1$ is the mobile-versus-cash payment contrast after cash remuneration. The corresponding contrast after mobile remuneration is $\beta_1 + \beta_3$. The interaction $\beta_3$ —the difference between the two conditional payment contrasts—is the focal factorial estimand in the present analysis. Adjusted cell means and simple contrasts are reported to interpret this interaction; their role is descriptive rather than to create additional primary hypotheses. The study and analysis plan were not preregistered. We therefore distinguish the factorial interaction implied by the complete design from secondary heterogeneity tests and exploratory decomposition and moderation.

For participant-level, within-wave permutation-based inference, we permute complete four-cell treatment labels within each implementation wave, preserve each wave's observed cell sizes, and keep both product observations attached to the participant. We re-estimate the preferred model in 9,999 permutations. This procedure asks how often an interaction at least as large in absolute value would arise under the maintained within-wave exchangeability of the rule-assigned treatment labels.

OLS is the primary estimator because its interaction coefficient directly expresses the difference between the two conditional payment contrasts in RMB. WTP is bounded at RMB 0 and RMB 20, but relatively few bids occur at either endpoint; two-limit Tobit and fractional-response models are reported as robustness checks. Secondary analyses examine whether the interaction differs across products or waves. Exploratory decompositions characterize whether the observed pattern resembles a symmetric same-medium premium or a directional one-cell departure. Robustness analyses vary the sample definition, aggregate WTP by participant, adjust for recorded pre-existing characteristics, and use alternative functional forms. Statistical significance is assessed at $p<.05$.

# 4. Results

## 4.1 Descriptive pattern

The primary sample contains 522 protocol-valid integer WTP observations from 274 participants. Table 2 reports the unadjusted cell distributions. Mean WTP is RMB 10.73 in the cash–cash condition, RMB 10.31 in cash–mobile, RMB 7.51 in mobile–cash, and RMB 9.69 in mobile–mobile. The unadjusted cell means do not indicate a uniform mobile-payment premium. Mean WTP is lowest in the mobile-remuneration/cash-payment condition.

**Table 2. Willingness to pay by receipt–payment path**

| Remuneration / payment | Participants | Product obs. | Mean | SD | Range |
|---|---|---|---|---|---|
| Cash / Cash | 77 | 142 | 10.73 | 4.44 | 1–20 |
| Cash / WeChat Pay | 58 | 114 | 10.31 | 4.75 | 0–20 |
| WeChat transfer / Cash | 71 | 141 | 7.51 | 4.53 | 0–20 |
| WeChat transfer / WeChat Pay | 68 | 125 | 9.69 | 4.27 | 0–20 |

*Notes.* Unadjusted summaries for protocol-valid integer WTP. Product observations are not exactly twice participant counts because every protocol-valid product response is retained. Participants valued a displayed mug and two RMB 10 McDonald's vouchers.

The rule-based allocation does not guarantee exact finite-sample balance. Participant-level diagnostics show the largest differences for cash-use and Alipay-use frequency; the remaining recorded characteristics are more closely balanced (Online Appendix Table A.2). The preferred design-based model does not condition on participant characteristics. A robustness model adjusts for sex, income category, phone-use time, and historical cash, Alipay, and WeChat Pay frequencies on an identical complete-covariate sample.

### 4.2 Primary factorial estimate

Table 3 reports the factorial regressions. The receipt-by-payment interaction is +RMB 2.60 in the unadjusted specification and remains +RMB 2.60 after product and implementation-wave fixed effects (SE = 0.88, 95% CI [0.86, 4.34], p=.004). This rejects the design-implied benchmark that the mobile-payment contrast is invariant to the medium of experimental remuneration.

**Table 3. Factorial regressions of willingness to pay**

| | **(1)** | **(2)** | **(3)** |
|---|---|---|---|
| Mobile purchase payment | -0.418<br>(0.676) | -0.400<br>(0.677) | -0.391<br>(0.677) |
| Mobile remuneration | -3.215***<br>(0.600) | -3.222***<br>(0.599) | -3.227***<br>(0.600) |
| Mobile payment × mobile remuneration | 2.596**<br>(0.884) | 2.610**<br>(0.883) | 2.603**<br>(0.884) |
| Constant | 10.725***<br>(0.438) | 11.795***<br>(0.457) | 11.863***<br>(0.523) |
| Product fixed effects | No | Yes | Yes |
| Implementation-wave fixed effects | No | No | Yes |
| Product observations | 522 | 522 | 522 |
| Participants (clusters) | 274 | 274 | 274 |
| R-squared | 0.075 | 0.128 | 0.128 |

*Notes.* OLS coefficients; participant-clustered standard errors in parentheses. Cash purchase payment and cash remuneration are the reference categories. Model 3 is preferred. * p < .05; ** p < .01; *** p < .001.

Figure 1 and Table 4 translate the preferred model into adjusted means and contrasts. After cash remuneration, adjusted WTP is RMB 10.72 with cash payment and RMB 10.33 with mobile payment. The mobile-versus-cash contrast is −RMB 0.39 (SE = 0.68, 95% CI [−1.72, 0.94], p=.564). After mobile remuneration, adjusted WTP is RMB 7.49 with cash payment and RMB 9.70 with mobile payment. The corresponding mobile-payment contrast is +RMB 2.21 (SE = 0.57, 95% CI [1.09, 3.33], p<.001). The difference between these two contrasts is the +RMB 2.60 interaction.

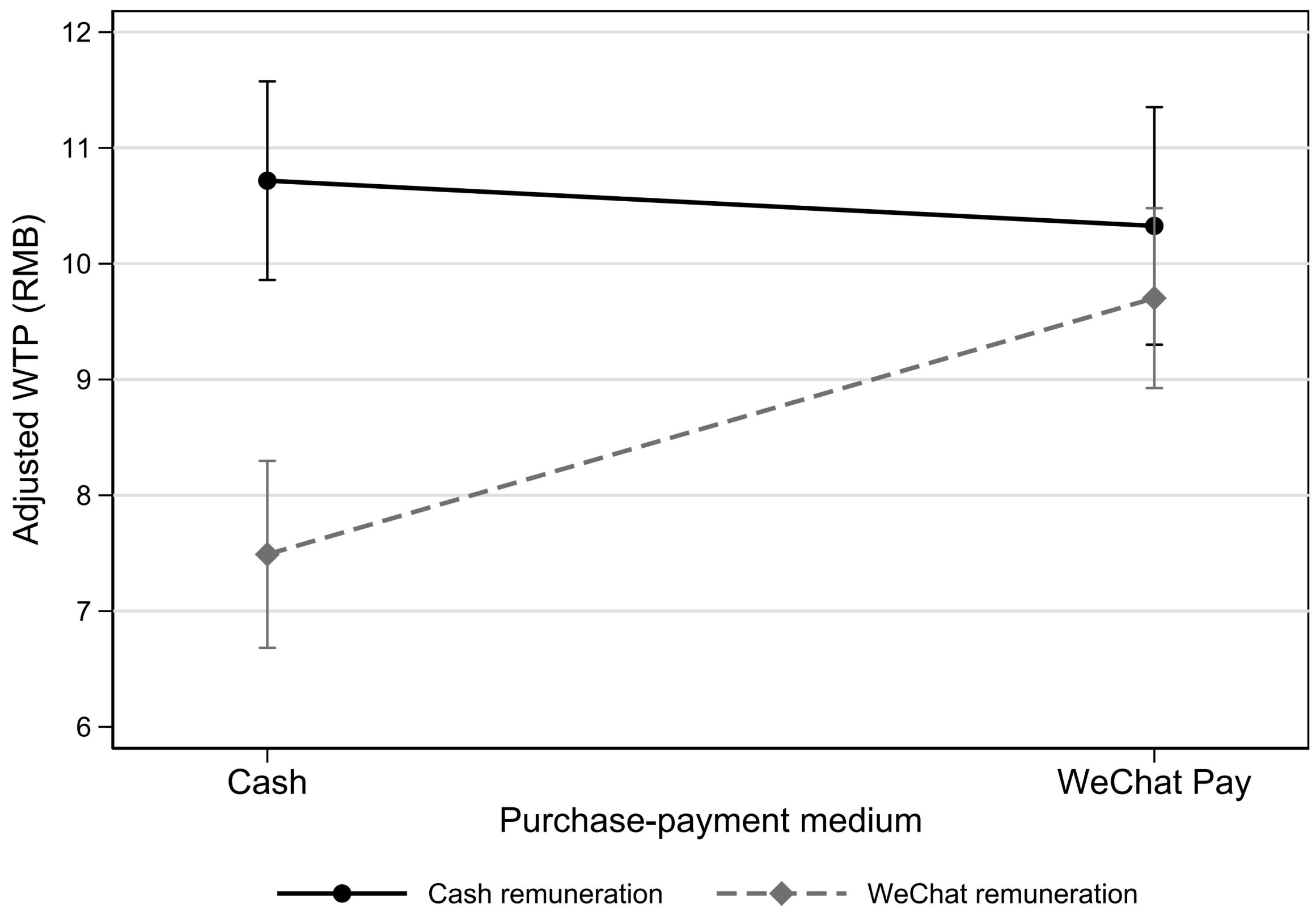

**Figure 1. Adjusted WTP in the 2 × 2 receipt–payment design**

**Table 4. Adjusted cell means and factorial contrasts**

**Panel A. Adjusted cell means**

| Receipt–payment path | Mean WTP | 95% CI |
|---|---|---|
| Cash remuneration / cash payment | 10.72 | [9.86, 11.57] |
| Cash remuneration / WeChat Pay | 10.33 | [9.30, 11.35] |
| WeChat transfer remuneration / cash payment | 7.49 | [6.68, 8.30] |
| WeChat transfer remuneration / WeChat Pay | 9.70 | [8.93, 10.48] |

**Panel B. Design-relevant contrasts**

| Estimand | Estimate | SE | 95% CI | p-value |
|---|---|---|---|---|
| Mobile-payment contrast after cash remuneration | -0.39 | 0.68 | [-1.72, 0.94] | .564 |
| Mobile-payment contrast after mobile remuneration | 2.21 | 0.57 | [1.09, 3.33] | <.001 |
| Remuneration-by-payment interaction | 2.60 | 0.88 | [0.86, 4.34] | .004 |
| Equal-weighted average mobile-payment contrast | 0.91 | 0.44 | [0.04, 1.78] | .040 |

*Notes.* Estimates from Model 3 in Table 3. Confidence intervals and p-values use participant-clustered standard errors. N = 522 product observations from 274 participants.

Equal-weighting the two remuneration conditions produces a pooled mobile-payment contrast of +RMB 0.91 (SE = 0.44, 95% CI [0.04, 1.78], p=.040). The pooled result is therefore positive but conceals materially different conditional contrasts. Relative to the residual dispersion in the preferred model, the interaction corresponds to 0.60 standard deviations (95% CI [0.20, 0.99]).

The participant-level, within-wave permutation test supports the same conclusion. Only 40 of 9,999 treatment-label permutations generate an interaction at least as large in absolute value as the observed RMB 2.60 estimate. With the conventional add-one correction, the two-sided permutation p-value is .004. Under the maintained within-wave exchangeability assumption, interactions of this magnitude are rare under reassignment of the observed treatment labels.

### *4.3 Exploratory characterization of the asymmetry*

The interaction is the focal factorial estimand in the present analysis. The following decomposition is exploratory and describes the form of the four-cell result. Equality of the cash–cash, cash–mobile, and mobile–mobile adjusted means cannot be rejected jointly (p=.221). By contrast, the restriction implied by a symmetric same-medium premium is rejected (p<.001). A parsimonious indicator for the mobile-remuneration/cash-payment cell estimates that this path lowers WTP by RMB 2.78 relative to the other three paths (SE = 0.49, p<.001).

The evidence therefore does not support a general rule that matching receipt and payment media raises valuation. It identifies a directional departure concentrated in the WeChat-to-cash path. In these data, an additive specification containing only PayMobile and SameMedium yields positive coefficients while masking the one-cell asymmetry underlying them.

### *4.4 Consistency across products and implementations*

Figure 2 reports the interaction for each product and implementation. The interaction is +RMB 2.34 for the mug (95% CI [0.34, 4.34], p=.022) and +RMB 2.86 for the voucher pair (95% CI [0.59, 5.14], p=.014). A formal product-heterogeneity test does not reject a common interaction (p=.672). The receipt–payment dependence is therefore observed both for a product without a stated price and for a product with a salient nominal face-value anchor.

The interaction is +RMB 3.05 in the first implementation (95% CI [0.55, 5.55], p=.017) and +RMB 2.20 in the second (95% CI [−0.22, 4.62], p=.075). The second estimate is less precise, but the point estimates have the same sign and a similar magnitude; a formal wave-heterogeneity test does not reject equality (p=.629). Because implementation, university, and date coincide, these results demonstrate internal consistency across two implementations rather than a separately identified campus or time effect.

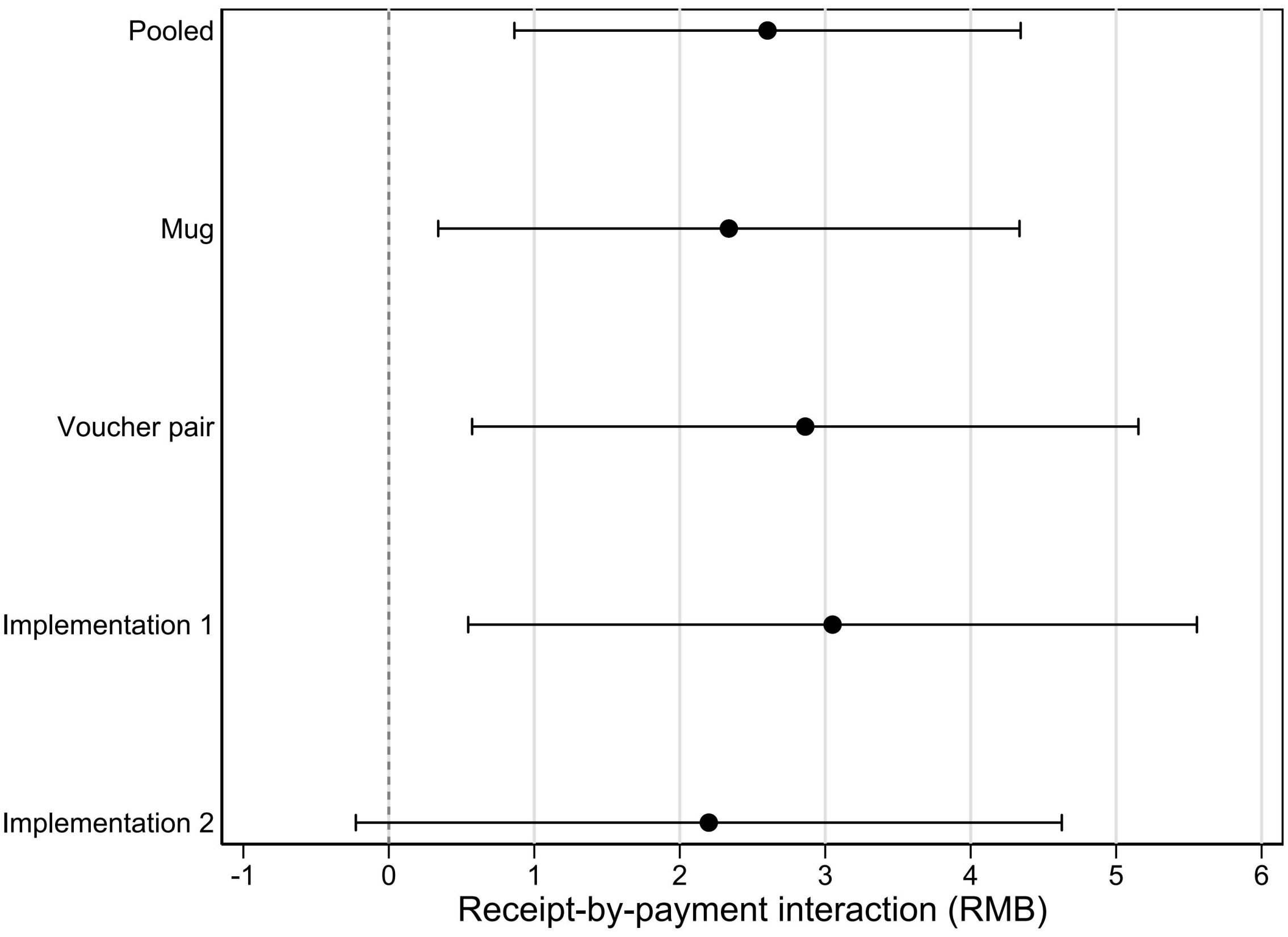

**Figure 2. Pooled, product-specific, and implementation-specific interactions**

The interaction is positive in all four product-by-wave subsamples. These finer estimates are necessarily imprecise and are reported in Online Appendix Figure A.1 as descriptive consistency checks, not as four additional hypothesis tests.

### *4.5 Robustness*

Table 5 summarizes the principal sample and aggregation checks. Requiring paired-complete integer bids yields an interaction of +RMB 2.47, averaging WTP to one observation per participant yields +RMB 2.68, excluding recorded comprehension failures yields +RMB 2.80, and retaining recorded comprehension passes only yields +RMB 3.19. Across the sample-definition and aggregation checks reported in Table 5, the estimates are positive and statistically significant. A deliberately permissive specification that retains all 531 nonmissing bids, preserves decimals, and caps the five out-of-range bids at the protocol bounds estimates an interaction of +RMB 2.42 (SE = 0.90, 95% CI [0.65, 4.18], p=.008). The result therefore does not depend on treating the protocol-valid integer sample as the sole admissible definition. Because protocol-valid response rates differed across cells for the mug, we also assigned every non-protocol-valid product record to the outcome bound least favorable to the positive interaction. Under this deliberately extreme scenario, the interaction remained positive at RMB 0.61, although its 95% confidence interval included zero; the opposite endpoint assignment yielded RMB 4.48 (Online Appendix Table A.5).

**Table 5. Robustness to sample definitions and aggregation**

| Specification | Interaction | SE | 95% CI | Obs. | Participants |
|---|---|---|---|---|---|
| Protocol-valid integer bids (primary) | 2.60** | 0.88 | [0.86, 4.34] | 522 | 274 |
| Paired protocol-valid participants | 2.47** | 0.91 | [0.68, 4.26] | 496 | 248 |
| Participant-mean WTP | 2.68** | 0.91 | [0.89, 4.46] | 274 | 274 |
| No recorded comprehension failure | 2.80** | 0.92 | [0.98, 4.62] | 485 | 254 |
| Recorded comprehension pass only | 3.19** | 1.00 | [1.23, 5.16] | 390 | 204 |
| All nonmissing bids, capped to [0,20] | 2.42** | 0.90 | [0.65, 4.18] | 531 | 275 |

*Notes.* Models include product and implementation-wave fixed effects where defined. Standard errors are clustered by participant except in the participant-mean model. The final row retains all nonmissing bids, including in-range decimals, and caps out-of-range bids at RMB 0 and RMB 20. * $p < .05$; ** $p < .01$; *** $p < .001$.

On the complete-covariate sample, the preferred model without participant characteristics estimates an interaction of +RMB 2.66; adding sex, income category, phone-use time, and historical cash, Alipay, and WeChat Pay frequencies produces +RMB 2.72. This is descriptive covariate adjustment, not an alternative source of identification, but it indicates that the interaction is not attenuated by adjustment for the recorded pre-existing characteristics.

Alternative functional forms produce the same qualitative conclusion. The interaction is positive in log(WTP+1) OLS, a participant random-intercept model, a two-limit Tobit model, and fractional logit for WTP/20. The complete estimates appear in Online Appendix Table A.6; covariate-adjustment checks appear in Table A.7. The robustness analysis is used to evaluate dependence on reasonable modeling choices, not to select the most favorable specification.

### *4.6 Exploratory questionnaire evidence*

The post-valuation questionnaire does not identify a single psychological mediator. Neither the adapted tightwad–spendthrift proxy nor mobile-payment familiarity, perceived relative payment pain, price attention, spending-control perceptions, convenience, or the Wave 2 financial-monitoring scale provides evidence of systematic moderation of the factorial interaction. The common specification and complete estimates are reported in Online Appendix Table A.8.

Because the attitudinal measures were collected after WTP and do not measure transaction-state processes, they are supplementary rather than mediational evidence. The central result remains the design-based receipt-medium × payment-medium interaction.

## 5. General Discussion

### *5.1 The mobile-payment effect begins before checkout*

Payment research usually locates the treatment at the moment money leaves. Our experiment shows that a salient monetary event immediately before checkout can change the contrast observed there. Mobile rather than cash payment produces little estimated difference after cash remuneration but a sizeable positive difference after WeChat remuneration. The factorial interaction is approximately RMB 2.60, equivalent to 0.60 residual standard deviations of WTP. It is supported by participant-level permutation inference and remains robust across alternative samples, participant aggregation, covariate adjustment, and functional forms.

This is more specific than the familiar conclusion that context matters. The complete crossing of receipt and payment media identifies the context: the route through which immediately available funds enter the budget changes the contrast attributed to the route through which they leave. A pooled mobile-payment coefficient therefore averages over behaviorally different monetary paths. This observation offers a new way to understand why cashless effects can be positive on average yet vary markedly across designs and settings.

### 5.2 Remuneration is part of the payment environment

The first contribution is methodological. Incentive protocols do not merely determine whether choices are consequential; they can define the funds against which those choices are made. Recent work shows that the incidence and probability of experimental payment can affect elicited preferences (Aydogan et al., 2024; Berlin et al., 2026). Our results extend this insight to payment form. An equal participation amount delivered as banknotes or through a mobile interface does not necessarily create the same subsequent cash–mobile contrast.

This distinction completes a design left open by earlier payment experiments. Runnemark et al. (2015) did not implement account remuneration followed by cash purchase because cash availability could not be assured. Liu et al. (2021) varied money source while aligning the medium used to provide and spend it. By ensuring that participants could transact in both media and independently varying receipt and purchase media, our design estimates the previously unobserved factorial interaction.

The implication is practical and substantive. When researchers provide spendable funds shortly before a consequential choice, the remuneration medium belongs in the description of the payment environment. A cash-versus-mobile comparison following newly supplied cash and the same comparison following a digital transfer answer different empirical questions. The protocol can therefore alter both the magnitude and the interpretation of the "payment effect."

### 5.3 Why the WeChat-to-cash path is distinctive

The second contribution is behavioral. The four treatment cells do not reveal a general same-medium premium. Instead, the interaction is concentrated in lower valuation when an immediately received WeChat transfer is followed by a required cash purchase. We call this receipt–payment path dependence: the value assigned to a product varies with the sequence linking a salient recent inflow to the medium of the ensuing outflow.

Incomplete behavioral fungibility provides one interpretation of this pattern. Newly received money may be technically available while remaining associated with a particular account, interface, or anticipated use. A WeChat transfer is represented in a digital environment. When purchase also occurs through WeChat Pay, receipt and payment share an interface and action system, although the data do not identify which linked funding source ultimately settles the transaction. When purchase instead requires cash, the digital gain remains visible while the consumer must part with physical money already held. The new inflow and the required outflow can therefore remain behaviorally separated even though they belong to the same aggregate budget.

Cash remuneration need not create the mirror image. Newly received notes can be physically commingled with cash already held and can make the cash budget salient before either cash or mobile checkout. Cash also remains broadly usable across payment environments, whereas moving a digital balance into physical cash involves a more explicit change of form. The mobile-to-cash condition uniquely combines a salient digital gain, a cross-medium action change, and the loss salience of relinquishing banknotes. Medium-linked mental accounting, account salience, cross-medium friction, and physical-cash loss salience are therefore complementary explanations of the asymmetry. For payment experiments, this asymmetry has a direct design implication: when a recently assigned digital inflow is followed by a required cash outflow, the cross-medium transition becomes part of the treatment environment rather than a neutral logistical detail.

The experiment identifies the path dependence rather than a unique state mechanism. That distinction opens a focused research agenda. Future studies can vary whether balances are displayed, whether funds can be converted within the session, and the delay between receipt and purchase; they can also measure transaction-state account integration and payment discomfort. The present contribution is the empirical restriction those mechanisms must explain: the same checkout contrast changes with the preceding receipt medium, and the change is directional rather than symmetrically matching.

### 5.4 Consistency across products and implementations

The interaction is similar for a mug without a stated price and for vouchers with a combined nominal face value of RMB 20. The voucher pair does not impose a common subjective valuation, but it supplies an explicit monetary anchor. The persistence of the interaction across both products shows that receipt–payment path dependence is not confined to goods whose value lacks a salient reference.

The interaction also has the same sign and a comparable magnitude in two implementations conducted six months apart at different universities. The second estimate is less precise, but formal heterogeneity tests do not distinguish the two implementation effects. Combined with the product results and specification checks, this cross-implementation consistency makes the factorial pattern more informative than a result tied to a single good, session, or estimator.

### 5.5 Broader relevance and historical scope

The experiment isolates a recently received participation payment, but the underlying sequence is common outside the laboratory. Consumers receive transfers, refunds, rebates, gifts, bonuses, and platform credits into identifiable payment environments before making later purchases. These inflows differ in origin, salience, restrictions, and interface. Our results suggest that their effects cannot be understood only from their amount: the route of receipt can shape how the next payment medium affects valuation.

The 2018–2019 Chinese setting offers an unusually informative dual-medium window. WeChat Pay was already routine, while cash remained familiar and liquid enough for every receipt–payment path to be behaviorally natural. In a more cashless environment, requiring cash may itself create exceptional friction; in an earlier setting, mobile payment might have been unfamiliar. The study therefore supplies a historical baseline from a period in which physical and digital money coexisted as everyday media rather than treating one as an artificial alternative.

The evidence is drawn from university participants making low-value purchases, and the experiment does not directly manipulate the proposed state processes. These boundaries define where extension is most valuable, but they do not diminish the central factorial finding that receipt medium changes the payment contrast.

### *5.6 Conclusion*

Experimental money has a path as well as an amount. In this study, the contrast between mobile and cash payment changed with the medium through which participation money had just been received. This result broadens the relevant payment environment: remuneration is not merely a logistical prelude to a consequential purchase but can shape the effect attributed to checkout itself. More broadly, consumer valuation can be path dependent: it reflects not only the payment friction at checkout but also the sequence through which money enters and subsequently leaves the decision environment.

[1] North China Electric Power University, Information Coding Standard, student-number section (the final six positions form the serial number within category and enrollment year), https://its.ncepu.edu.cn/gzzd/b3d42e4c5648464694addba95d8cb2e7.htm; China University of Petroleum (Beijing), Public Information Coding Standard, section 4.3 (personnel codes use YYYYCCNNNN, where NNNN is an administratively allocated serial number), https://www.cup.edu.cn/nic/zczd/xjzd/850436c0ab2c48048ce5cf5562d661bf.htm .

# References

Abeler, J., & Marklein, F. (2017). Fungibility, Labels, and Consumption. Journal of the European Economic Association, 15(1), 99–127. https://doi.org/10.1093/jeea/jvw007

Agarwal, S., Ghosh, P., Li, J., & Ruan, T. (2024). Digital Payments and Consumption: Evidence from the 2016 Demonetization in India. The Review of Financial Studies, 37(8), 2550–2585. https://doi.org/10.1093/rfs/hhae005

Arkes, H. R., Joyner, C. A., Pezzo, M. V., Nash, J. G., Siegel-Jacobs, K., & Stone, E. (1994). The Psychology of Windfall Gains. Organizational Behavior and Human Decision Processes, 59(3), 331–347. https://doi.org/10.1006/obhd.1994.1063

Aycinena, D., Bogliacino, F., & Kimbrough, E. O. (2026). Replication: Dictator Games with Different Endowment Sources. Journal of Economic Psychology, 116, 102921. https://doi.org/10.1016/j.joep.2026.102921

Aydogan, I., Berger, L., & Théroude, V. (2024). Pay All Subjects or Pay Only Some? An Experiment on Decision-Making under Risk and Ambiguity. Journal of Economic Psychology, 104, 102757. https://doi.org/10.1016/j.joep.2024.102757

Becker, G. M., DeGroot, M. H., & Marschak, J. (1964). Measuring Utility by a Single-Response Sequential Method. Behavioral Science, 9(3), 226–232. https://doi.org/10.1002/bs.3830090304

Berlin, N., Kemel, E., Lenglin, V., & Nebout, A. (2026). Paying None, Some or All? Between-Subject Random Incentives and Preferences towards Risk and Time. Journal of Economic Psychology, 112, 102870. https://doi.org/10.1016/j.joep.2025.102870

Broekhoff, M. C., & van der Cruijsen, C. (2024). Paying in a Blink of an Eye: It Hurts Less, but You Spend More. Journal of Economic Behavior & Organization, 221, 110–133. https://doi.org/10.1016/j.jebo.2024.03.017

Carver, C. S., & White, T. L. (1994). Behavioral Inhibition, Behavioral Activation, and Affective Responses to Impending Reward and Punishment: The BIS/BAS Scales. Journal of Personality and Social Psychology, 67(2), 319–333. https://doi.org/10.1037/0022-3514.67.2.319

Clot, S., Grolleau, G., & Ibanez, L. (2018). Shall we pay all? An experimental test of random incentivized systems. Journal of Behavioral and Experimental Economics, 73, 93–98. https://doi.org/10.1016/j.socec.2018.01.004

Feinberg, R. A. (1986). Credit Cards as Spending Facilitating Stimuli: A Conditioning Interpretation. Journal of Consumer Research, 13(3), 348–356. https://doi.org/10.1086/209074

Gelman, M., & Roussanov, N. (2024). Managing Mental Accounts: Payment Cards and Consumption Expenditures. The Review of Financial Studies, 37(8), 2586–2624. https://doi.org/10.1093/rfs/hhae013

Hastings, J. S., & Shapiro, J. M. (2013). Fungibility and Consumer Choice: Evidence from Commodity Price Shocks. The Quarterly Journal of Economics, 128(4), 1449–1498. https://doi.org/10.1093/qje/qjt018

Heath, C., & Soll, J. B. (1996). Mental Budgeting and Consumer Decisions. Journal of Consumer Research, 23(1), 40–52. https://doi.org/10.1086/209465

Henderson, P. W., & Peterson, R. A. (1992). Mental Accounting and Categorization. Organizational Behavior and Human Decision Processes, 51(1), 92–117. https://doi.org/10.1016/0749-5978(92)90006-S

Lee, J. N., Morduch, J., Ravindran, S., & Shonchoy, A. S. (2024). The Social Meaning of Mobile Money: Earmarking Reduces the Willingness to Spend in Migrant Households. Journal of Economic Behavior & Organization, 221, 675–688. https://doi.org/10.1016/j.jebo.2024.04.023

Levav, J., & McGraw, A. P. (2009). Emotional Accounting: How Feelings about Money Influence Consumer Choice. Journal of Marketing Research, 46(1), 66–80. https://doi.org/10.1509/jmkr.46.1.66

Liu, Y., & Dewitte, S. (2021). A Replication Study of the Credit Card Effect on Spending Behavior and an Extension to Mobile Payments. Journal of Retailing and Consumer Services, 60, 102472. https://doi.org/10.1016/j.jretconser.2021.102472

Liu, Y., Luo, J., & Zhang, L. (2021). The Effects of Mobile Payment on Consumer Behavior. Journal of Consumer Behaviour, 20(3), 512–520. https://doi.org/10.1002/cb.1880

Ma, Q., Tan, Y., He, Y., Cheng, L., & Wang, M. (2024). Why Does Mobile Payment Promote Purchases? Revisiting the Pain of Paying, and Understanding the Implicit Pleasure via Selective Attention. PsyCh Journal, 13(5), 760–779. https://doi.org/10.1002/pchj.765

Milkman, K. L., & Beshears, J. (2009). Mental Accounting and Small Windfalls: Evidence from an Online Grocer. Journal of Economic Behavior & Organization, 71(2), 384–394. https://doi.org/10.1016/j.jebo.2009.04.007

Prelec, D., & Loewenstein, G. (1998). The Red and the Black: Mental Accounting of Savings and Debt. Marketing Science, 17(1), 4–28. https://doi.org/10.1287/mksc.17.1.4

Prelec, D., & Simester, D. (2001). Always Leave Home Without It: A Further Investigation of the Credit-Card Effect on Willingness to Pay. Marketing Letters, 12(1), 5–12. https://doi.org/10.1023/A:1008196717017

Raghubir, P., & Srivastava, J. (2008). Monopoly Money: The Effect of Payment Coupling and Form on Spending Behavior. Journal of Experimental Psychology: Applied, 14(3), 213–225. https://doi.org/10.1037/1076-898X.14.3.213

Reinholtz, N., Bartels, D. M., & Parker, J. R. (2015). On the Mental Accounting of Restricted-Use Funds: How Gift Cards Change What People Purchase. Journal of Consumer Research, 42(4), 596–614. https://doi.org/10.1093/jcr/ucv045

Reshadi, F., & Fitzgerald, M. P. (2023). The Pain of Payment: A Review and Research Agenda. Psychology & Marketing, 40(8), 1672–1688. https://doi.org/10.1002/mar.21825

Rick, S. I., Cryder, C. E., & Loewenstein, G. (2008). Tightwads and Spendthrifts. Journal of Consumer Research, 34(6), 767–782. https://doi.org/10.1086/523874

Runnemark, E., Hedman, J., & Xiao, X. (2015). Do Consumers Pay More Using Debit Cards than Cash? Electronic Commerce Research and Applications, 14(5), 285–291. https://doi.org/10.1016/j.elerap.2015.03.002

Schomburgk, L., Belli, A., & Hoffmann, A. O. I. (2024). Less Cash, More Splash? A Meta-analysis on the Cashless Effect. Journal of Retailing, 100(3), 382–403. https://doi.org/10.1016/j.jretai.2024.05.003

Soman, D. (2001). Effects of Payment Mechanism on Spending Behavior: The Role of Rehearsal and Immediacy of Payments. Journal of Consumer Research, 27(4), 460–474. https://doi.org/10.1086/319621

Soman, D. (2003). The Effect of Payment Transparency on Consumption: Quasi-Experiments from the Field. Marketing Letters, 14(3), 173–183. https://doi.org/10.1023/A:1027444717586

Thaler, R. (1985). Mental Accounting and Consumer Choice. Marketing Science, 4(3), 199–214. https://doi.org/10.1287/mksc.4.3.199

Thaler, R. H., & Johnson, E. J. (1990). Gambling with the House Money and Trying to Break Even: The Effects of Prior Outcomes on Risky Choice. Management Science, 36(6), 643–660. https://doi.org/10.1287/mnsc.36.6.643

Thaler, R. H. (1999). Mental Accounting Matters. Journal of Behavioral Decision Making, 12(3), 183–206. https://doi.org/10.1002/(SICI)1099-0771(199909)12:3<183::AID-BDM318>3.0.CO;2-F

Voslinsky, A., & Azar, O. H. (2021). Incentives in experimental economics. Journal of Behavioral and Experimental Economics, 93, 101706. https://doi.org/10.1016/j.socec.2021.101706

# Online Appendix

How Money Enters Before It Leaves: Experimental Remuneration and the Mobile-Payment Effect

This appendix reports supplementary diagnostics, robustness analyses, and a faithful English rendering of the core participant-facing instructions. Redacted public copies of the original Chinese questionnaires and faithful English renderings accompany the replication package.

## Appendix A. Supplementary empirical results

### A.1 Sample flow

**Table A.1. Participant flow by implementation and treatment cell**

| Implementation | Treatment | Source records | Any valid WTP | Paired valid |
|---|---|---|---|---|
| Implementation 1 | WeChat remuneration / cash payment | 41 | 41 | 41 |
| Implementation 1 | WeChat remuneration / mobile payment | 40 | 40 | 31 |
| Implementation 1 | Cash remuneration / cash payment | 45 | 45 | 33 |
| Implementation 1 | Cash remuneration / mobile payment | 29 | 28 | 28 |
| Implementation 2 | WeChat remuneration / cash payment | 30 | 30 | 29 |
| Implementation 2 | WeChat remuneration / mobile payment | 28 | 28 | 26 |
| Implementation 2 | Cash remuneration / cash payment | 32 | 32 | 32 |
| Implementation 2 | Cash remuneration / mobile payment | 30 | 30 | 28 |

*Notes*. Any valid WTP identifies participants contributing at least one protocol-valid integer response. Paired valid requires valid responses for both products.

## A.2 Participant-level balance

### Table A.2. Recorded participant characteristics by treatment cell

| Characteristic | CC | CM | MC | MM | Joint p | N |
|---|---|---|---|---|---|---|
| Female | 0.43 | 0.47 | 0.36 | 0.56 | .113 | 272 |
| Income category | 3.99 | 4.00 | 4.17 | 4.06 | .897 | 271 |
| Daily phone-use category | 4.14 | 4.39 | 4.23 | 4.41 | .148 | 272 |
| Cash-use frequency | 2.18 | 1.97 | 2.00 | 1.99 | .024 | 274 |
| Alipay-use frequency | 4.01 | 3.71 | 4.24 | 4.15 | .035 | 274 |
| WeChat Pay-use frequency | 4.48 | 4.61 | 4.49 | 4.59 | .605 | 275 |

*Notes.* CC = cash remuneration/cash payment; CM = cash remuneration/mobile payment; MC = WeChat remuneration/cash payment; MM = WeChat remuneration/mobile payment. Joint p-values test treatment indicators conditional on implementation wave. These diagnostics describe finite-sample balance; they are not used to select covariates.

## A.3 Response validity and bid boundaries

### Table A.3. Protocol-valid response rate by product and treatment cell

| Product | CC | CM | MC | MM | Joint p |
|---|---|---|---|---|---|
| Mug | 92.2% | 94.9% | 100.0% | 89.7% | <.001 |
| McDonald's voucher pair | 92.2% | 98.3% | 98.6% | 94.1% | .169 |

*Notes.* Linear-probability omnibus tests include implementation-wave fixed effects and robust standard errors. The main analysis retains every protocol-valid product response; paired-complete estimates are reported as a robustness check.

### Table A.4. Distributional summary and protocol boundaries

| Product | Observations | WTP = 0 | WTP = 20 | Mean | SD |
|---|---|---|---|---|---|
| Mug | 259 | 2 (0.8%) | 12 (4.6%) | 10.59 | 4.26 |
| McDonald's voucher pair | 263 | 10 (3.8%) | 7 (2.7%) | 8.46 | 4.80 |

*Notes.* The limited endpoint mass motivates OLS as the primary estimator; bounded-outcome estimators are reported below.

### Table A.5. Extreme-value sensitivity to all non-protocol-valid product records

| Assignment scenario | Interaction | SE | 95% CI | p-value |
|---|---|---|---|---|
| Lower bound for positive interaction | 0.611 | 0.915 | [-1.190, 2.412] | .505 |
| Upper bound for positive interaction | 4.482 | 0.924 | [2.663, 6.300] | <.001 |

*Notes.* The analysis assigns every non-protocol-valid product record to the endpoint that minimizes or maximizes the positive factorial interaction, while retaining all 550 potential product records. These deliberately extreme scenarios bound the effect of outcome-dependent response validity; they are not estimates of plausible missing values.

## *A.4 Alternative functional forms*

### Table A.6. Alternative functional forms

| Model | Reported scale | Interaction | SE | 95% CI | Observations |
|---|---|---|---|---|---|
| Log(WTP + 1) OLS | Log points | 0.399*** | 0.107 | [0.188, 0.610] | 522 |
| Random-intercept model | WTP (RMB) | 2.638** | 0.874 | [0.925, 4.352] | 522 |
| Two-limit Tobit | Predicted WTP (RMB) | 2.636** | 0.895 | [0.882, 4.389] | 522 |
| Fractional logit | Predicted WTP (RMB) | 2.602** | 0.880 | [0.878, 4.326] | 522 |

*Notes.* All models include product and implementation-wave fixed effects. The log-OLS row reports the coefficient in log points. The mixed model reports the WTP interaction in RMB. For Tobit and fractional logit, the table reports the cross-difference in predicted observed WTP, converted to RMB. OLS standard errors are clustered by participant; the mixed model includes a participant random intercept. * $p < .05$; ** $p < .01$; *** $p < .001$.

## *A.5 Covariate adjustment*

### Table A.7. Descriptive covariate-adjustment robustness

| Specification | Interaction | SE | 95% CI | Obs. | Participants |
|---|---|---|---|---|---|
| Same complete-covariate sample, no control | 2.659** | 0.897 | [0.893, 4.426] | 512 | 269 |
| Same sample, recorded characteristics | 2.722** | 0.891 | [0.968, 4.476] | 512 | 269 |

*Notes.* The adjusted model includes sex, income category, daily phone-use category, and historical cash, Alipay, and WeChat Pay frequencies. Both rows use the identical complete-covariate sample.

### *A.6 Detailed consistency estimates*

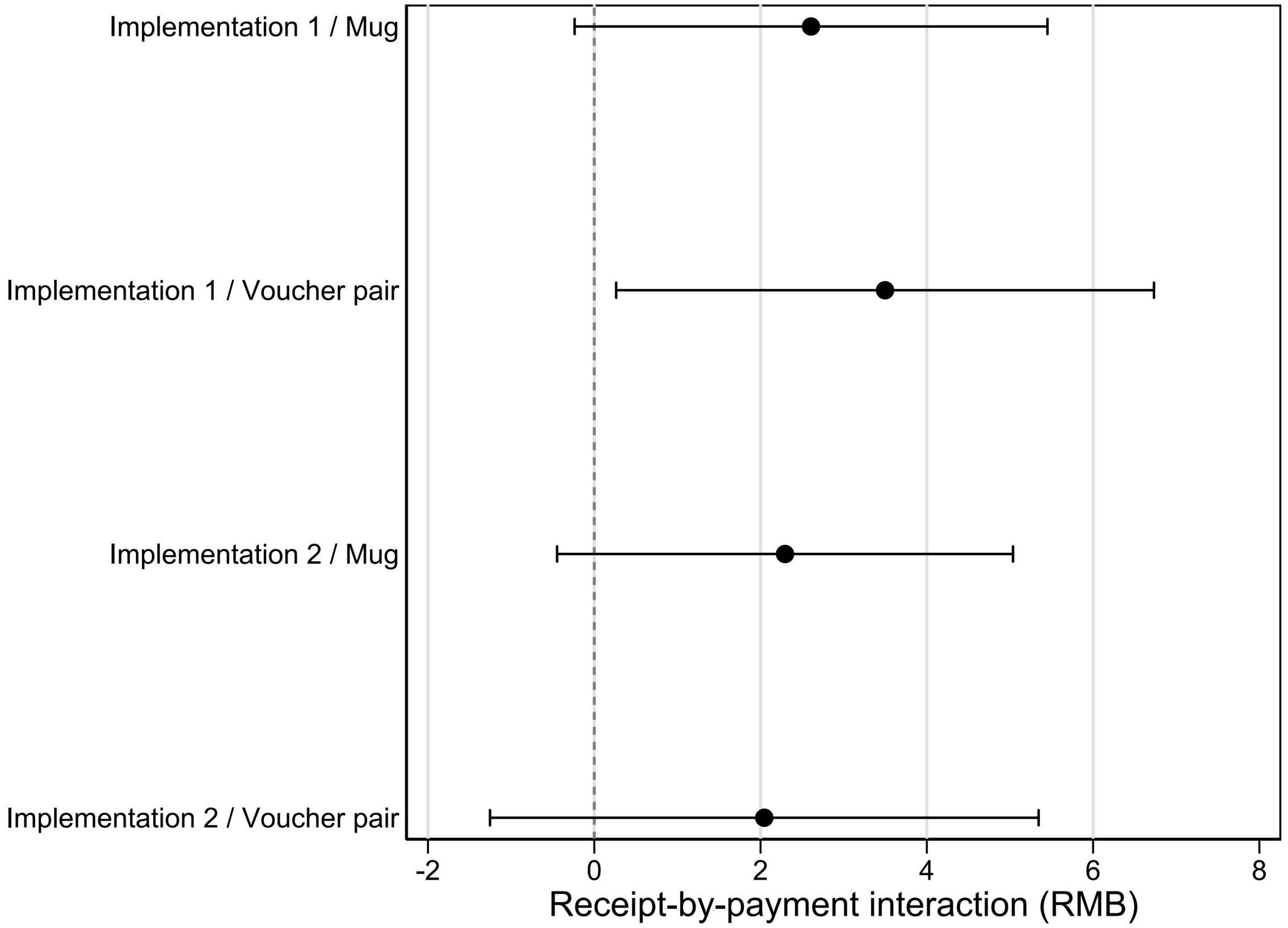

**Figure A.1. Receipt-by-payment interaction in each product × implementation subsample**

Unlike Figure 2, which reports pooled and one-dimensional product- or implementation-specific estimates, Figure A.1 jointly stratifies the sample by both product and implementation. The four product-by-implementation estimates are descriptive consistency checks. Their confidence intervals are wide because the factorial cells are divided into small subsamples; the formal heterogeneity tests are reported in the main text.

### *A.7 Exploratory moderation*

**Table A.8. Exploratory moderation of the factorial interaction**

| Moderator | Triple interaction | SE | p | Joint p | Participants |
|---|---|---|---|---|---|
| Trait pain sensitivity | -0.092 | 0.876 | .916 | .963 | 257 |
| Mobile-payment familiarity | 0.058 | 0.953 | .952 | .234 | 274 |
| Perceived relative payment pain | 0.796 | 0.954 | .405 | .683 | 273 |
| Price attention | 0.097 | 0.857 | .910 | .913 | 273 |
| Spending-control perceptions | 0.199 | 0.801 | .804 | .374 | 274 |
| Perceived convenience | 0.391 | 0.750 | .602 | .610 | 274 |
| Financial monitoring | 0.996 | 1.153 | .389 | .242 | 120 |

*Notes.* Each moderator is standardized. Joint p-values test all moderator-related interaction terms. Trait pain sensitivity reverses the adapted spending-orientation score so that higher values indicate greater difficulty spending. Mobile-payment familiarity averages standardized WeChat, Alipay, relative mobile-use, and reverse-coded cash-use frequencies. The remaining constructs use the corresponding questionnaire items described in the replication code. Measures were collected after WTP and are not interpreted as mediators.

## Appendix B. Participant-facing materials

### *B.1 Consent procedure*

Before the experiment, participants received the Chinese participant-information and consent form approved for the study and provided informed consent before taking part. No completed forms or participant-identifying information are shared.

### *B.2 Core purchasing instructions (faithful English rendering)*

You have an opportunity to purchase Product A or Product B. We will collect each participant's willingness to pay (WTP) and determine the outcome as follows.

- Step 1. Participants are assigned with equal probability to the Product A group or the Product B group based on the experimental questionnaire ID shown below.
- Step 2. For each product, an integer from RMB 0 to RMB 20 will be selected with equal probability as the product price.

- Step 3. We will compare the participant's WTP with the selected product price. If WTP is equal to or greater than the product price, the participant will purchase the product at the selected price using the payment method specified in that questionnaire version. If WTP is below the product price, no purchase will occur.

WTP must be an integer from RMB 0 to RMB 20.

Analogy used in the instructions: imagine asking a friend to buy an item whose price you do not know. You tell the friend to buy it if the price is no more than RMB XX (your WTP), and otherwise not to buy it. The actual price is independent of your WTP; your WTP determines only whether a purchase occurs.

The questionnaire then provided two numerical examples and one comprehension question. Participants were invited to raise their hands if anything was unclear.

Participants reported: (1) WTP for the mug, in RMB; and (2) WTP for the pair of McDonald's vouchers, in RMB.

Implementation note (not questionnaire wording). Participants submitted both WTP bids before learning which product the parity of their experimental questionnaire ID selected. The selected product and random price were disclosed only after both bids had been returned. Researchers reviewed the examples and explained the comprehension answer before bidding.

### B.3 Treatment-specific wording

**Table B.1. Treatment-specific receipt and purchase wording**

| Implementation | Condition | Remuneration and preparation instruction | Payment |
|---|---|---|---|
| Implementation 1 | Cash–cash | Receive RMB 30 cash; place the RMB 30 cash on the desk | Cash |
| Implementation 1 | Cash–mobile | Receive RMB 30 cash; open the mobile-payment application | WeChat Pay |
| Implementation 1 | Mobile–cash | Receive RMB 30 by WeChat; place the previously requested RMB 30 cash on the desk | Cash |
| Implementation 1 | Mobile–mobile | Receive RMB 30 by WeChat; open the mobile-payment application | WeChat Pay |
| Implementation 2 | Cash–cash | Receive RMB 30 cash; place the required RMB 20 cash on the desk | Cash |
| Implementation 2 | Cash–mobile | Receive RMB 30 cash; open the mobile-payment application | WeChat Pay |
| Implementation 2 | Mobile–cash | Receive RMB 30 by WeChat; place the required RMB 20 cash on the desk | Cash |
| Implementation 2 | Mobile–mobile | Receive RMB 30 by WeChat; open the mobile-payment application | WeChat Pay |

*Notes.* This table preserves the substantive differences across all eight original Chinese questionnaires. WeChat was used for both mobile remuneration and mobile purchase payment.

### B.4 Questionnaire modules

After WTP, both implementations measured an adapted tightwad–spendthrift orientation, past-year frequencies of cash, debit-card, credit-card, Alipay, and WeChat Pay use, relative cash/mobile preference and use, eight perceptions of cash and mobile payment, nonbinding intended quantities, purchase outcome, a purchase-contingent repurchase WTA, phone use, income category, and sex. The first implementation also contained a tablet-choice vignette. The second implementation additionally included financial-monitoring items and the BIS/BAS instrument. Mobile-account and purchase-funding questions varied by implementation and treatment version. The translated item inventory and the variable-construction code accompany the replication package. Redacted public Chinese reference questionnaires, faithful English renderings for both implementations, and a treatment-wording crosswalk covering all eight versions are also included. The intended quantities, WTA, tablet choice, and account questions are disclosed but are not manuscript outcomes. The remaining post-valuation measures are used only in exploratory analyses.